\documentclass[twocolumn,10pt,letterpaper]{article}

\usepackage{amsmath}
\usepackage{tikz}
\usepackage{url}
\usepackage{subfigure}
\usepackage{tabularx}
\usepackage{placeins}
\usetikzlibrary{intersections}
\usetikzlibrary{angles,quotes}

\def\D{\mathrm{d}}
\def\vrms{v_\mathrm{RMS}}
\def\eqperiod{\text{.}}
\def\eqcomma{\text{,}}

\newcommand{\myfigurewide}[3]
{\begin{figure*}[btp]\begin{center}#2\caption{#3}\label{#1}\end{center}\end{figure*}}

\newcommand{\myfigure}[3]
{\begin{figure}[btp]\begin{center}#2\caption{#3}\label{#1}\end{center}\end{figure}}

\newcommand{\mygraphfigure}[4]{\myfigure{#1}{\includegraphics[#2]{#3}}{#4}}

\newcommand{\mygraphfigurewide}[4]{\myfigurewide{#1}{\includegraphics[#2]{#3}}{#4}}

\newcommand{\mysubfigure}[3]
{\subfigure[#3]{\hfill#2\label{#1}\hfill}}

\newcommand{\mygraphsubfigure}[4]
{\mysubfigure{#1}{\includegraphics[#2]{#3}}{#4}}

\newcommand{\mytable}[3]
{\begin{table}[btp]\begin{center}#2\caption{#3}\label{#1}\end{center}\end{table}}

\begin{document}

\title{A Review of ERA5 Equatorial Wind Data \protect\\ and Impact on a Space Elevator}

\author{Blaise Gassend\thanks{gassend@alum.mit.edu} \and Jonathan
Gassend}
\date{}


\maketitle

\begin{abstract}

  This paper uses the ERA5 weather dataset to assess the impact of the wind on an
  equatorial space elevator. An analytical formulation establishes that
  the effect of wind on tether inclination and payload capacity can be
  captured by a single scalar: RMS wind velocity, averaged over air pressure.
  This formulation clarifies what constitutes excessive wind for a
  particular elevator and provides a direct link between wind data and
  elevator design parameters.

  Analysis of equatorial ERA5 data shows that equatorial wind loading
  varies strongly with geography with lower values over continents and
  higher values over oceans, with the majority contribution coming
  from the lowest 20~km of atmosphere. The data indicate that realistic equatorial winds
  pose a significant constraint on thin-ribbon elevator designs if align to
  face the wind.  Combining wind and lightning statistics highlights
  several promising equatorial regions for anchor sites. 

  Finally, the implications of wind
  loading for elevator design are examined, and several mitigation
  strategies are discussed to improve survivability and operational payload
  capacity under realistic wind conditions.

\end{abstract}


\section*{List of Symbols}

\begin{tabularx}{\columnwidth}{l X}
  $\epsilon$ & Ribbon tether thickness. (m)\\
  $\theta$ & Angle between tether and vertical. Equivalently, angle between
             the wind and the tether normal. (rad or deg) \\
  $\rho_a$ & Density of air. ($\mathrm{kg/m^3}$) \\

\end{tabularx}
\begin{tabularx}{\columnwidth}{l X}

  $\rho_t$ & Density of tether. ($\mathrm{kg/m^3}$) \\
  $\mu$ & Dynamic viscosity of air. ($\mathrm{Pa\cdot s}$) \\
  $\sigma$ & Stress in tether. ($\mathrm{N/m^2}$)\\
  $A$ & Tether cross-section $A=\epsilon W$. ($\mathrm{m^2}$) \\
  $F_\perp$ & Normal component of wind force on the tether. ($\mathrm{N/m}$)  \\
  $C_D$ & Drag coefficient of the tether ribbon. (flat 2D plate) \\
  $F_\parallel$ & Tangential component of wind force on the tether,
                  oriented upward along the tether. ($\mathrm{N/m}$) \\
  $g$ & Acceleration due to gravity. ($\mathrm{m/s^2}$) \\
  $h$ & Height above the ground. (m) \\
  $P, P_0$ & Atmospheric pressure, at ground with 0 suffix. ($\mathrm{N/m^2}$) \\
  $R$ & Reynolds number. \\
  $s$ & Curvilinear coordinate increasing towards the ground. (m) \\
  $T$ & Tension in the tether. (N) \\
  $v$ & Wind velocity. (m/s) \\
  $\hat{v}$ & Non-dimensional wind velocity. \\
  $\vrms$ & RMS wind velocity averaged over pressure. (m/s) \\
  $v_c$ & Characteristic wind velocity. (m/s) \\
  $W$ & Width of the tether. (m) \\
\end{tabularx}

\section{Introduction}

A space elevator is a structure that extends from the surface of the Earth,
or some other celestial body, past the geostationary orbit and ends in a
counterweight, which maintains the structure in tension. First proposed by
Tsiolkovsky~\cite{tsiolkovsky1895} in 1895, and brought to attention in
the West by Pearson~\cite{pearson75}, the basis for the present space elevator
architecture is by Edwards~\cite{edwards2002}, where an elevator made of a thin
ribbon of carbon nanotubes or graphene is ascended by climbers powered by
lasers or other means.

Susceptibility of a space elevator to wind depends roughly on the ratio of
width to cross-section, since wind load will be proportional to width, and
strength will be proportional to cross-section. Pearson~\cite{pearson75} 
considered wind loading in his paper, but dismissed it because his
three 10~cm radius cables offer ample strength to resist wind loads.
Because of its thin ribbon shape, the Edwards elevator is far more prone
to wind loading. In~\cite{edwards2002}, Edwards places
the base of his elevator near the Galapagos Islands citing low
surface wind data.

More recent work by Robinson and Knapman~\cite{robinson2022} 
considered some aggregate altitude-dependent wind data, and found that the
wind loading on the space elevator is a substantial concern. They then
proposed several mitigation approaches which involve changing how payloads
get out of the atmosphere, some of them particularly challenging to
implement such as Lofstrom loops~\cite{lofstrom1985launch}, multi-staged
elevators~\cite{knapman2019multistage}, Thoth
towers~\cite{quine2015spaceelevator} or transferring from a winch to a
climbable elevator outside the atmosphere.


In this paper we extend the work from~\cite{robinson2022} by refining the criteria 
for determining whether wind loading is excessive, and using the ERA5
database to explore in more depth the wind distributions in the equatorial
region. We also suggest the capstan drive as a path for allowing the
narrower ribbons required for wind loading survivability to be climbed,
avoiding the need to fundamentally change the structure of the elevator
within the atmosphere. Finally, we point out that~\cite{robinson2022} used
a value of the drag coefficient that is too low.

The paper begins with an analysis section that finds the effect of wind on
inclination and payload capacity of the elevator in
Section~\ref{sec:analysis}. Then in Section~\ref{sec:data} it presents
findings mined from the ERA5 wind database about wind conditions in the
equatorial region. Finally, Section~\ref{sec:design} shows the impact of the
wind on elevator design and proposes mitigations.

\section{Analysis}
\label{sec:analysis}

\subsection{Shape Equation for a Tether}

\myfigure{wind-tether}{
  \begin{tikzpicture}
  \coordinate (P) at (1,2);

  \draw[->, thick] (3.5, 2) node[above left] {wind} -- (2.5, 2);

  \draw[->, thick] (-1,0) -- (-1,5.5) node[above] {$h$};
  
  \draw[ultra thick, name path=ribbon, ->] (0,5) node[above] {ribbon} to[out=-85,in=135] (P) to [out=-45, in=155] (4,0) node[right] {$s$};

  \draw[->,red,thick] (P) -- +(-45:2) node[below left] {$\vec{T}$};

  \draw[->,blue,thick] (P) -- +(-135:1.5) node[below] {$\vec{e_\perp}$};

  \draw[->,blue,thick] (P) -- +(135:1.5) node[above left] {$\vec{e_\parallel}$} coordinate (Q);

  \draw[->,brown,thick] (P) -- +(-90:1) node[below left] {$\vec{G}$};
  
  \draw[->,brown,thick] (P) -- +(-160:1.5) node[below left] {$\vec{W}$};

  \draw[thick] (P) -- +(0, 1.5cm);
  \path (Q) coordinate (C) -- (P) coordinate (B) -- +(0,1) coordinate (A) pic [draw, ->, "$\theta$", angle radius=0.8cm] {angle};

\end{tikzpicture}

}{Forces acting on the tether.}

First, we consider the equilibrium of a tether under tension T in the
presence of load $\vec{F}$ per unit length, and decompose into a basis
$(\vec{e_\parallel}, \vec{e_\perp})$ formed with vectors locally parallel and normal
to the tether according to the convention in Figure~\ref{wind-tether}.

\begin{equation}
\begin{aligned}
  \vec{F} & = -\frac{\D \vec{T}}{\D s}
   = \frac{\D T \vec{e_\parallel}}{\D s} 
   = T\frac{\D \vec{e_\parallel}}{\D s} + \frac{\D T}{\D s} \vec{e_\parallel} \\
   & = T\frac{\D \theta}{\D s} \vec{e_\perp} + \frac{\D T}{\D s} \vec{e_\parallel} 
   \eqcomma
\end{aligned}
\end{equation}

\noindent where $\theta$ is the inclination of the tether at curvilinear
coordinate $s$, where $s$ increases down the elevator.

Rearranging yields a classic equation for the equilibrium shape of a tether subjected
to a force
\begin{align}
  \frac{\D \theta}{\D s} & = \frac{F_\perp}{T} \\
  \frac{\D T}{\D s} & = F_\parallel 
  \eqcomma
\end{align}
\noindent where the first equation shows how the normal component of the
applied force sets the curvature of the tether, and the tangential
component causes variation in tension.

Now we can use the fact that $$\frac{\D h}{\D s} = -\cos \theta$$ to get
rid of $s$, yielding
\begin{align}
  \label{eq:theta-eq}
  \frac{\D \theta}{\D h} & = - \frac{F_\perp}{T \cos \theta} \\
  \label{eq:t-eq}
  \frac{\D T}{\D h} & = - \frac{F_\parallel}{\cos \theta}
  \eqperiod
\end{align}

The force on the tether is due to a combination of gravity and wind:
$\vec{F} = \vec{G} + \vec{W}$. Taking the ratio of $T$ over $G$, we can see
that $G$ has little impact within the distance scale of the atmosphere:

\begin{equation}
\begin{aligned}
  \label{eq:gravity-small}
  \frac{T}{G} & = \frac{\sigma A}{\rho_t A g} = \frac{\sigma}{\rho_t g} 
  \approx \frac{10\ \mathrm{GPa}}{2300\ \mathrm{kg\ m^3} \times 9.8\
  \mathrm{m/s^2}} \\
  & = 440\ \mathrm{km}
  \eqcomma
\end{aligned}
\end{equation}

\noindent where $\sigma$, $\rho_t$ and $A$ are the stress, density and
cross-section for the tether, and $g$ is the acceleration of gravity at
the surface of the Earth. We assumed that the tether was graphene and that
it was lightly loaded at about 10~\% of its capacity.

Equation~\ref{eq:gravity-small} shows that gravity will have a minor effect
on the tether shape within a few tens of km of atmosphere, even when the
tether is unloaded, except once the tether inclination becomes high. We
will neglect the effect of gravity for the remainder of this paper, and
focus our attention on the wind forces.

Equation~\ref{eq:t-eq} shows that the wind will cause an increase in
tension near the base of the tether making the tether less susceptible to
the wind in Equation~\ref{eq:theta-eq}. Unless large inclinations are
reached, this increase will also be modest. We will therefore conservatively
assume uniform tension. It would be possible to evaluate the inaccuracy
introduced by this approximation on solutions, but we will not consider
that here.

\subsection{Wind Forces}

In \cite{taylor1952}, a model is proposed for the forces applied by a fluid
on a long smooth flexible cylinder. This model is considered accurate as long as
the angle between the flow and the axis of the cylinder is at least
20\textdegree{}. In this model, the cross-flow principle is applied, in which
the flow is assumed to decompose into a normal and a tangential component,
which each act independently on the cylinder. The normal component is
further divided into a component made of forces normal to the surface,
which is independent of Reynolds number $R=Wv\rho_a/\mu$, and a component
made up of forces tangential to the surface which scales like $R^{-1/2}$,
where $W$ is the width of the cylinder, $v$ is the wind speed, and $\rho_a$ and
$\mu$ are the density and dynamic viscosity of the fluid.
The model is given by

\begin{align}
  \label{eq:wind-perp}
  F_\perp & = \frac{1}{2} \rho_a v^2 W 
    \left(C_D \cos^2 \theta + 4 R^\frac{1}{2} \cos^\frac{3}{2} \theta \right) \\
  F_\parallel & = \frac{1}{2} \rho_a v^2 W 
    \left( 5.4 R^\frac{1}{2} \sin \theta \cos^{\frac{1}{2}} \theta \right)
  \eqperiod
\end{align}

We will assume that this model holds in the case of a ribbon, except that
$C_D$ and the two numeric constants $4$ and $5.4$ would have to be changed.

As in~\cite{robinson2022}, we shall make the conservative assumption that
the elevator ribbon is aligned such that it faces the wind. Because the
wind will cause the ribbon's centerline to bend, it is plausible that the
elevator will spontaneously align in this direction, but a proper
aeroelastic study needs to be done to determine if this is really the case.
Under this assumption, the elevator ribbon looks like a 2D flat plate with
$C_D$ around 1.98~\cite{hoerner1965}. This is higher than the value of 1.28
used in~\cite{robinson2022}, which is for a square plate rather than a
highly elongated rectangular plate.

For our purposes, the Reynolds number is somewhat large. For a skinny 1 cm ribbon
and a very low 1 m/s wind, it is 405. Thus $4 R^\frac{1}{2} = 0.198$, which is
10~\% of $C_D$. At higher wind levels or for wider ribbons,
this fraction will diminish, and as long as $\theta$ doesn't approach
90\textdegree{} we will have an acceptable approximation to
Equation~\ref{eq:wind-perp} if we just ignore the $R^\frac{1}{2}$ term, or
lump it into $C_D$:

\begin{equation}
  \label{eq:wind-perp-simple}
  F_\perp \approx \frac{1}{2} \rho_a v^2 W C_D \cos^2 \theta
  \eqperiod
\end{equation}

For an edge-on ribbon, the situation would be reversed, and the $R^{1/2}$
term would be dominant.
                  
\subsection{Tether Inclination at the Ground and Characteristic Speed}

Combining equations~\ref{eq:theta-eq} and~\ref{eq:wind-perp-simple} gives
us an approximate equation for the shape of the wind loaded elevator

\begin{equation}
  \frac{\D \theta}{\D h} = -\frac{\rho_a v^2 W C_D \cos \theta}{2 T} 
  \eqcomma
\end{equation}
\noindent where $\rho_a$ is the atmospheric density.

We simplify, separate variables and integrate from the ground to
infinity:

\begin{equation}
  \int_0^{\theta_0}\frac{\D \theta}{\cos \theta} = 
    \int_0^\infty \frac{\rho_a v^2 W C_D \D h}{2 T}
    \eqperiod
\end{equation}

Carrying out the integrations and using the fact that $\frac{\D P}{\D h} = -\rho_a g$, we get:

\begin{equation}
  \ln\Bigg(\tan\bigg( \frac{\theta}{2}+\frac{\pi}{4}\bigg)\Bigg) = 
    \frac{W C_D P_0}{2 Tg} \frac{1}{P_0} \int_0^{P_0} v^2 \D P
  \eqcomma
\end{equation}
\noindent where $P_0$ is the pressure at ground level.

This expression can be rewritten as:

\begin{align}
  \theta_0 & = f\Bigg(\frac{\vrms}{v_c}\Bigg) \\
  f(\hat{v}) & = 2 \tan^{-1} e^{\hat{v}^2} - \frac{\pi}{2} \approx \hat{v}^2
  \quad\mathrm{(\hat{v}\ small)} \\
  v_c & =\sqrt{\frac{2 T g}{W C_D P_0}}=\sqrt{\frac{2\sigma \epsilon g}{C_D P_0}} \\
  v_{\mathrm{RMS}}&=\sqrt{\left\langle v\right\rangle^2_P}
     =\sqrt{\frac{1}{P_0} \int_0^{P_0} v^2 \D P}
   \eqcomma
\end{align}

\noindent where we see that the inclination $\theta_0$ at the base of the
elevator is given by applying a generic function $f$, plotted in
Figure~\ref{fig:inclination}, to a non-dimensional velocity
$\hat{v}=\frac{\vrms}{v_c}$. The non-dimensional velocity is formed by taking
$\vrms$, the root mean square (RMS) wind velocity averaged over pressure,
divided by a characteristic velocity $v_c$. The second expression for $v_c$
is insightful because $\sigma$ is set by the tether material
properties. So the only parameter available to adjust elevator
susceptibility to wind is the thickness $\epsilon$.

\mygraphfigure{fig:inclination}{width=\columnwidth}{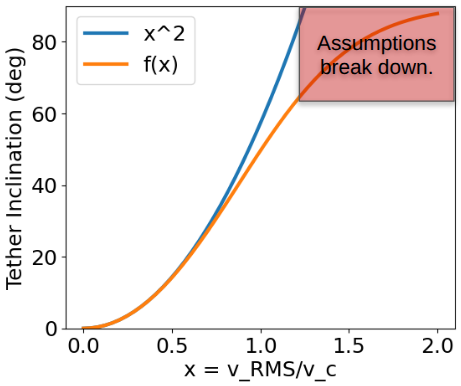}{Inclination
at the base of the space elevator as a function of non-dimensional wind speed.}

The information about wind
distribution all goes into $\vrms$, and information about the Earth and
elevator geometry go into $v_c$. This decoupling is convenient as it allows
us to compute $\vrms$ from weather data without making any assumptions
about the elevator, and then draw conclusions on acceptable values of
$v_c$, which can then lead to constraints on the elevator design.

To get a sense of the magnitudes, the International Space Elevator Consortium (ISEC) 
has an unloaded characteristic velocity of
\begin{equation}
  v_c =\sqrt{\frac{2\cdot90\ \mathrm{GPa}\cdot10\ \mathrm{\mu m}\cdot9.81\ 
      \mathrm{m/s^2}}{1.98\cdot10^5\ \mathrm{Pa}}}
      =9.4\ \mathrm{m/s}
  \eqperiod
\end{equation}

Below the climber the stress is reduced and the characteristic velocity is much lower. For instance, if we assume 10~\% tension below the tether we get:
\begin{equation}
  v_c  =\sqrt{\frac{2 \cdot 10\ \mathrm{GPa} \cdot 10\ \mathrm{\mu m} \cdot
    9.81\ \mathrm{m/s^2}} {1.98 \cdot 10^5\ \mathrm{Pa}}} = 3.1\ \mathrm{m/s}
  \eqperiod
\end{equation}

\subsection{Pull-Down}

As $\vrms$ increases substantially beyond $v_c$, the inclination at the
base of the elevator approaches 90\textdegree{}. The exact behavior in this
region is not within the domain of validity of our modeling as the
cross-flow principle breaks down. Moreover, large horizontal displacements
will occur, invalidating the flat-earth model implicit in our derivations.

We can nevertheless get an idea of what will happen in the high wind case.
The same analysis we have done at ground level also applies at
higher altitudes, and likewise concludes that 90\textdegree{} will be
approached at higher altitude for high wind. When this happens, there will be a large
stretch of ribbon that is near-horizontal between the higher altitude and
the ground, causing a large horizontal offset to exist between the anchor
and the location at which the elevator leaves the atmosphere.

If nothing is done at the anchor (e.g. feeding out ribbon), this horizontal
offset will cause the counterweight to be lowered. The tension caused by
the counterweight will therefore decrease. This decrease will cause $v_c$
to decrease, increasing the inclination, and causing more horizontal
offset. The result is a runaway condition where the elevator is blown
sideways as in Figure~\ref{fig:pull-down} and pulled down. This phenomenon
has been observed in numerical simulation \cite{lang2005}. One should note
that pull-down is not an immediate phenomenon and that a sustained
excessive wind will be needed to pull down the elevator.

\mygraphfigure{fig:pull-down}{width=\columnwidth}{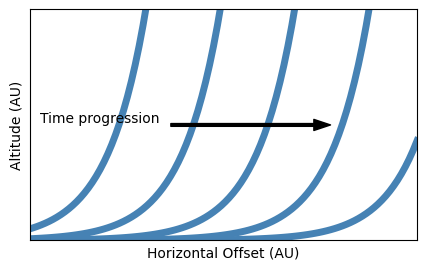}{Illustration
of the runaway condition that occurs for sufficient $\vrms$ causing the
elevator to be blown to the side resulting in it being pulled down.}

We will therefore set the requirement that at all times the inclination of
the ribbon at the anchor should be far enough from 90\textdegree{} to keep
us well away from the pull-down condition. Setting a limit such as 70\textdegree{} or
80\textdegree{} at the ground should be sufficient for this purpose.

\subsection{Load Fraction}

While large inclinations can be reached at the anchor, these inclinations
will compromise the ability to launch large payloads. Indeed, recall that
the wind inclines the tether without changing its tension much. Hence as the
tether inclines, the payload capacity gets scaled by $\cos \theta$.
Figure~\ref{fig:load-fraction} shows the remaining fraction of load
capacity as a function of $\vrms$, and Table~\ref{tab:v-over-vc} shows a
few specific values, along with the corresponding inclinations.
                                                
\mygraphfigure{fig:load-fraction}{width=\columnwidth}{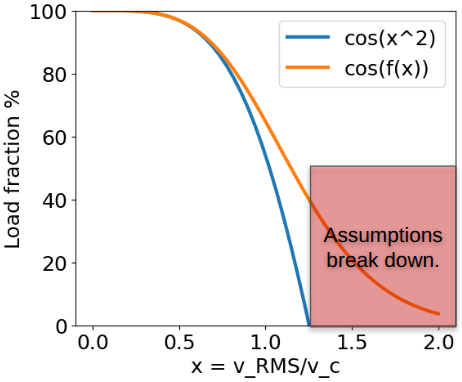}{Fraction
of load capacity that is available as a function of non-dimensional wind speed.}

\mytable{tab:v-over-vc}{
  \begin{tabular}{c | c | c}
  $v/v_c$ & Inclination (\textdegree{}) & Load Fraction (\%) \\
    \hline
    \hline
  0.56 & 18 & 95 \\
  0.68 & 26 & 90 \\
  0.83 & 37 & 80 \\
  1.15 & 60 & 50 \\
  1.32 & 70 & 34 \\
  1.56 & 80 & 17 \\
  \end{tabular}
}{Load fraction and anchor inclination for a few values of $\frac{v}{v_c}$.}

\section{Analysis of Geographic Data}
\label{sec:data}

\subsection{ERA5 Wind Data}

The European Centre for Medium-Range Weather Forecasts (ECMWF) compiles
the ECMWF ReAnalysis version 5 (ERA5) dataset \cite{hersbach2020era5,
cds_era5_hourly}. It provides the full state of the atmosphere extrapolated
by combining a computer model with available weather data. It covers the
time range from 1940 to present with a resolution of 1~h and 0.25\textdegree{} or
about 28~km at the equator. It includes 37 pressure levels from 1 hPa (about 60~km
altitude) to 1000 hPa. It provides several data products, and in particular
vertical and horizontal wind velocity.

Except where indicated, analyses in this paper were done using a sample covering
the whole ERA5 dataset from January 1940 to July 2025 with 3~h sampling. For
the equator, this dataset occupies about 100 GB.

Figure~\ref{fig:wind-equator-percentiles} shows the distribution of
$\vrms$ along the equator, showing that winds are lowest over the
continents at 16~m/s worst case and 12~m/s at the 90\textsuperscript{th} percentile. The
winds increase with distance from land reaching the highest speeds mid-Pacific
at 30~m/s worst case or 17~m/s at the 90\textsuperscript{th} percentile. It would be
informative to see whether climate models predict how these wind speed
distributions will change in the face of global warming.

\mygraphfigurewide{fig:wind-equator-percentiles}{width=0.9\textwidth}
{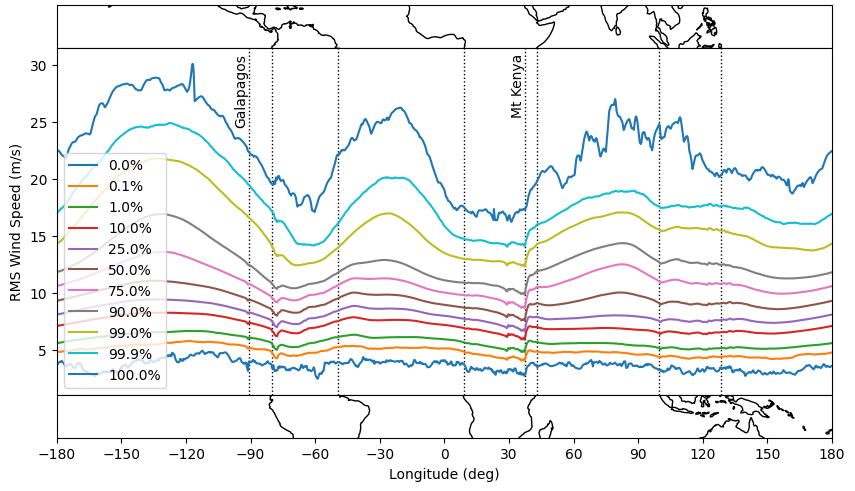}{ERA5 $\vrms$ percentiles as a function of
longitude, from 1940 to July 2025}

These numbers appear high compared with the 5~m/s wind speeds
in~\cite{edwards2002} at a location 100 miles West of the Galapagos
islands, which is about 105\textdegree{} West. Plotting the wind speed as a
function of altitude in Figure~\ref{fig:galapagos} explains the
discrepancy. The data from~\cite{edwards2002} is taken at ground level, but
the wind speed is much higher at 15~km altitude, and these higher altitudes
dominate the elevator wind loading.

\myfigurewide{fig:galapagos}{
  \mygraphsubfigure{fig:galapagos-v}{width=\columnwidth}{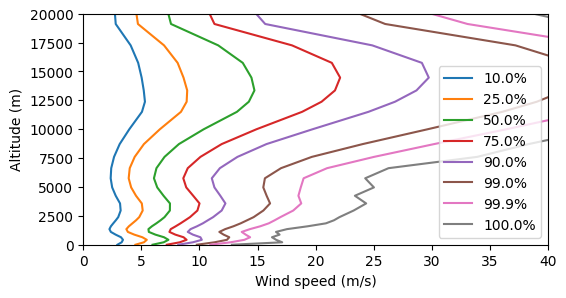}{Wind
  speed vs. altitude.}
  \mygraphsubfigure{fig:galapagos-pv2}{width=\columnwidth}{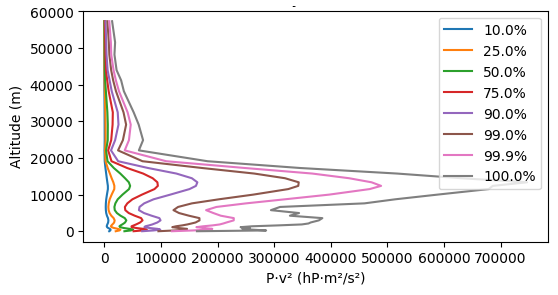}{$P v^2$ 
  vs. altitude.}
}{Wind speed and $P v^2$ (proportional to dynamic pressure) at
105\textdegree{} West, a location
roughly 1000 miles West of the Galapagos islands. The wind speed at the ground is
consistent with~\cite{edwards2002}, but the elevator wind loading is
dominated by higher altitudes.}

Figure~\ref{fig:seasonal} shows the seasonal variations of $\vrms$. The
Pacific Ocean has far stronger wind levels in the winter, whereas the
Indian Ocean has the strongest levels in the summer. The Atlantic Ocean
sees modestly higher winds both in the winter and summer.

\myfigurewide{fig:seasonal}{
  \mygraphsubfigure{fig:seasonal-max}{width=\columnwidth}{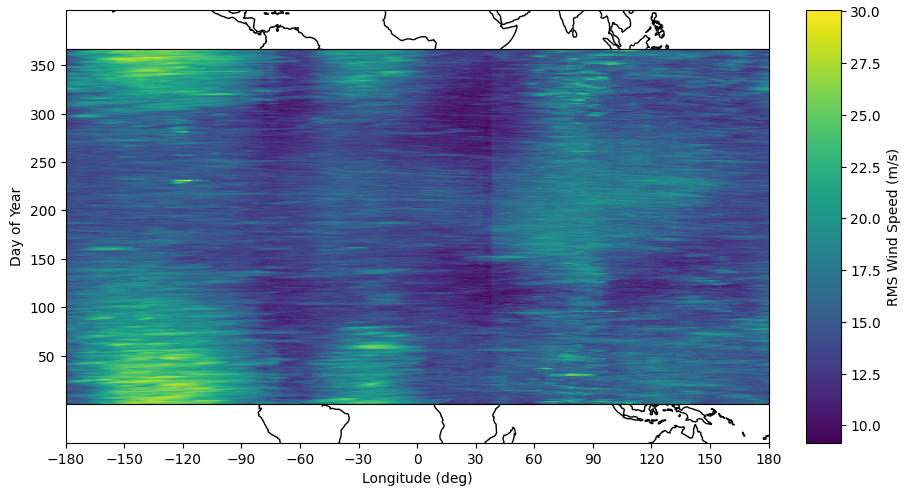}{
    Maximum $\vrms$ speed. 
  }
  \mygraphsubfigure{fig:seasonal-mean}{width=\columnwidth}{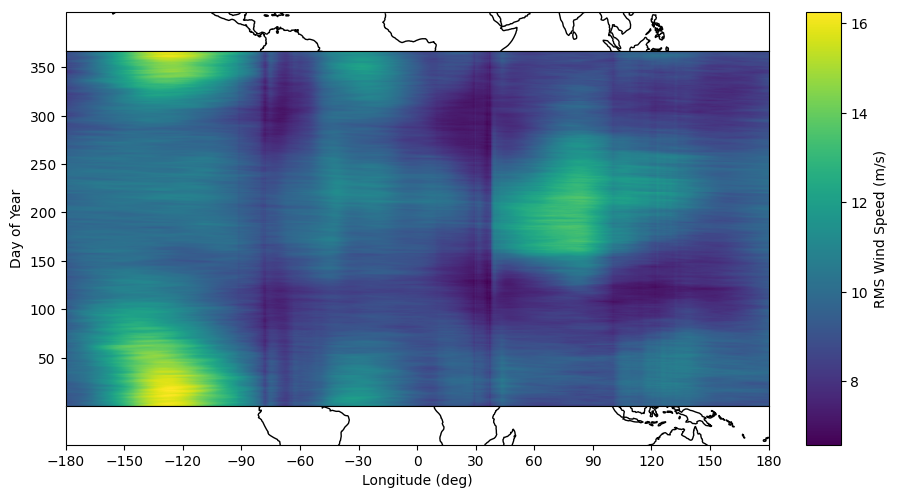}{
    Mean $\vrms$ speed.
  }
}{Seasonal variation of $\vrms$, from 1940 to July 2025.}

Figure~\ref{fig:wind-vs-years} shows clear variability in $\vrms$ from year
to year.  For instance, in February of 1983 the wind conditions in the
Equatorial Atlantic reached much higher levels than usual. Digging into
this case shows that a storm over North America caused a disruption to the
subtropical jet stream, which ended up extending down to the equatorial
region.

\mygraphfigure{fig:wind-vs-years}{width=\columnwidth}{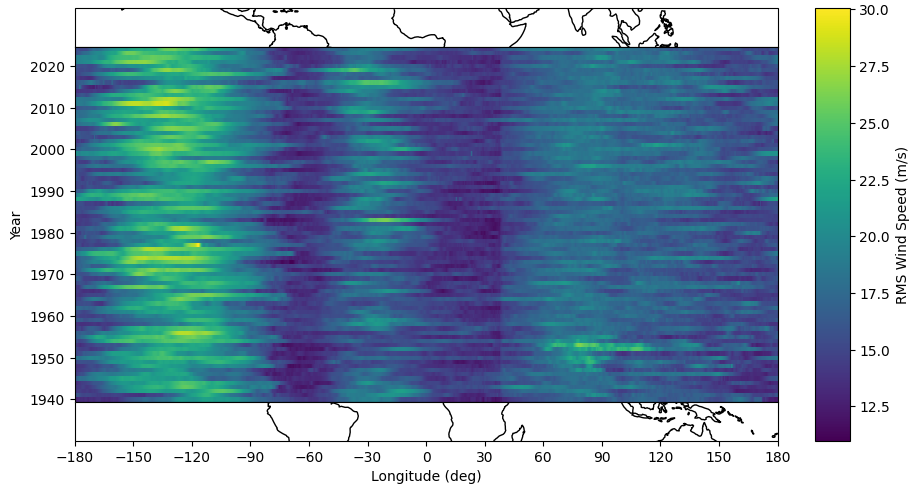}{Variation
of Maximum $\vrms$ from 1940 to July 2025, showing a steady background with
clear outlier events.}

Since it is possible for an elevator to be away from the
equator~\cite{gassend2004nonequatorial}, we also plot some non-equatorial
$\vrms$ results in Figure~\ref{fig:vrms-non-equatorial}. Because this is a
much larger data set, we only obtained daily samples for 2024. With modest
excursion away from the equator, the trends we have observed for the wind
at the equator still hold.

\mygraphfigurewide{fig:vrms-non-equatorial}{width=0.9\textwidth}
{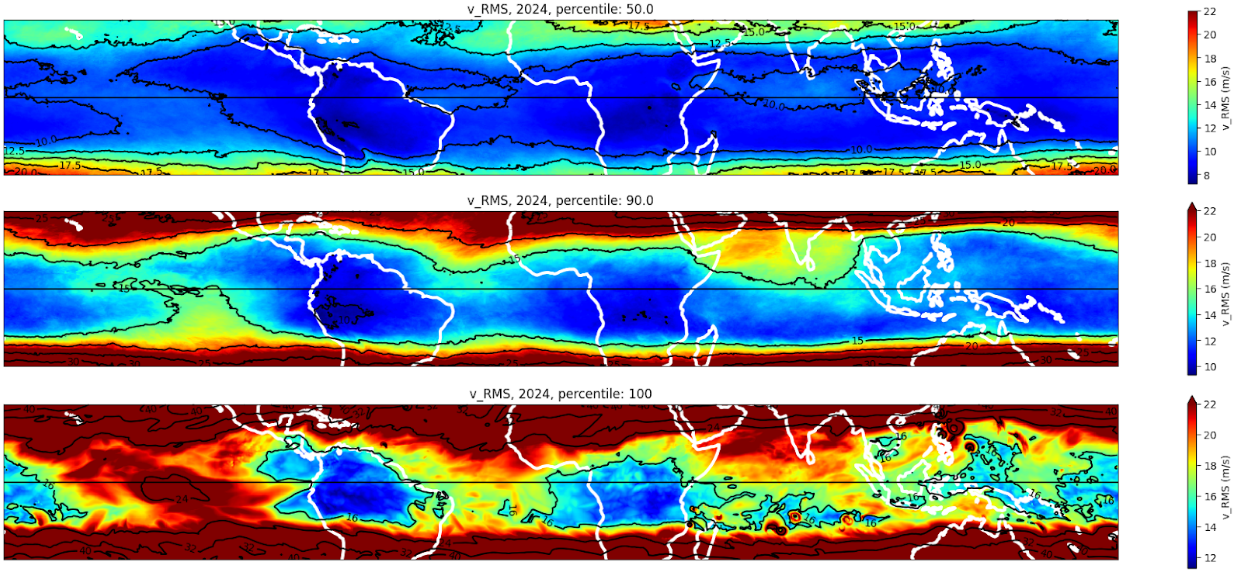}{ERA5 $\vrms$ 50\textsuperscript{th},
90\textsuperscript{th} and 100\textsuperscript{th} percentiles for 2024 up
to 25\textdegree{} of latitude.} 

Figure~\ref{fig:vertical-wind} shows the vertical component\footnote{The
vertical velocity component was converted from Pa/s to m/s assuming a
characteristic height of pressure decay of 8.3~km.  Given the small
magnitude of the wind speeds that were observed, a more accurate
computation was not deemed useful.} of the wind aggregated over 2024. The
vertical component is consistently small, usually below 0.5~m/s.  Higher
values can be seen below 20~km, peaking between 2.5 and 3~m/s.  Looking at
an instantaneous cross-section of the atmosphere in
Figure~\ref{fig:vertical-vs-alt-long}, it appears that the updrafts are
very localized. These updrafts are most likely tropical deep convection.

By Equations~\ref{eq:theta-eq} and~\ref{eq:t-eq}, updrafts
will tend to reduce the inclination of the elevator and increase the
tension at its base, with downdrafts having the opposite effect. But 
the observed wind speeds are so low that these effects are most likely
irrelevant.

\myfigurewide{fig:vertical-wind}{
  \mygraphsubfigure{fig:vertical-vs-altitude}{width=\columnwidth}{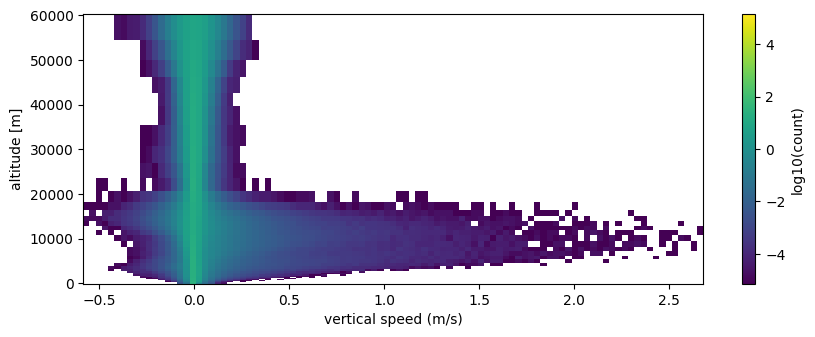}{
    Distribution of vertical wind component vs. altitude.
  }
\mygraphsubfigure{fig:horizontal-vs-altitude}{width=\columnwidth}{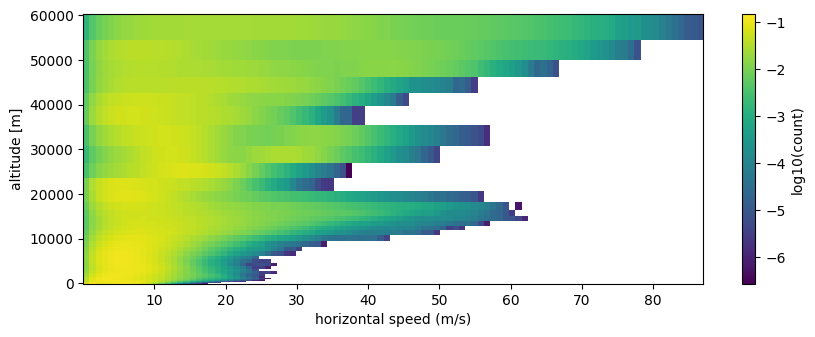}{
    Distribution of horizontal wind component vs. altitude.
  }
  \mygraphsubfigure{fig:vertical-vs-longitude}{width=\columnwidth}{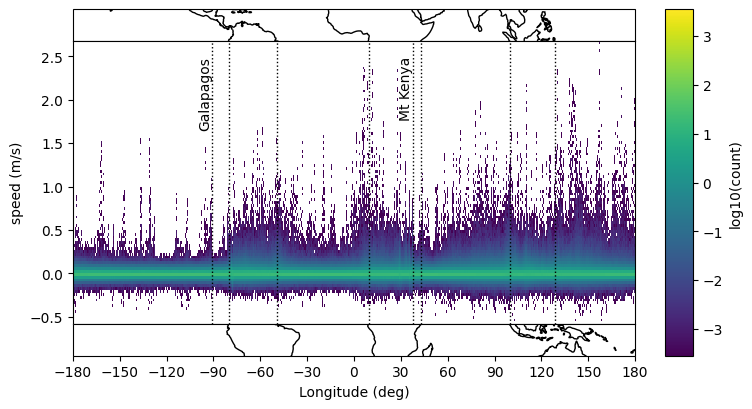}{
    Distribution of vertical wind component vs. longitude.
  }
  \mygraphsubfigure{fig:vertical-vs-alt-long}{width=\columnwidth}
  {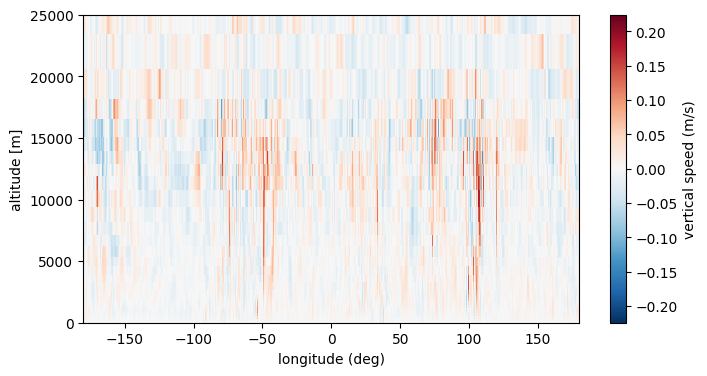}{
    Distribution of vertical wind component vs. altitude and longitude for 1/1/2024.
  }
}{
  Plots giving an idea of the vertical wind distribution compared with the
  horizontal distribution. These plots use a dataset limited to 2024.
  }

\FloatBarrier

\subsection{Low Lightning Locations}

In addition to low wind, it is also important to select an anchor location
that has a low rate of lightning strikes, as suggested in
\cite{edwards2002}. We now combine our results from the ERA5 database with
data from Vaisala Xweather~\cite{vaisala-lightning} downloaded in August
2025 to search for suitable anchor sites. If power beaming to the climbers
was going to be done using laser beaming, a region with low cloud cover
would also be needed. We will not further consider that power-beaming
requirement in this paper.

As seen in Figure~\ref{fig:lightning}, a few locations stand out combining
the low winds near the continents and low lightning rates: east and
west of equatorial South America, Suriname or adjacent regions
in Brazil, the equatorial region south of Liberia, north of Mount Kenya,
off the eastern equatorial coast of Africa, and northeast of Papua New
Guinea.

\myfigure{fig:lightning}{
  \mygraphsubfigure{fig:lightning-america}{width=\columnwidth}{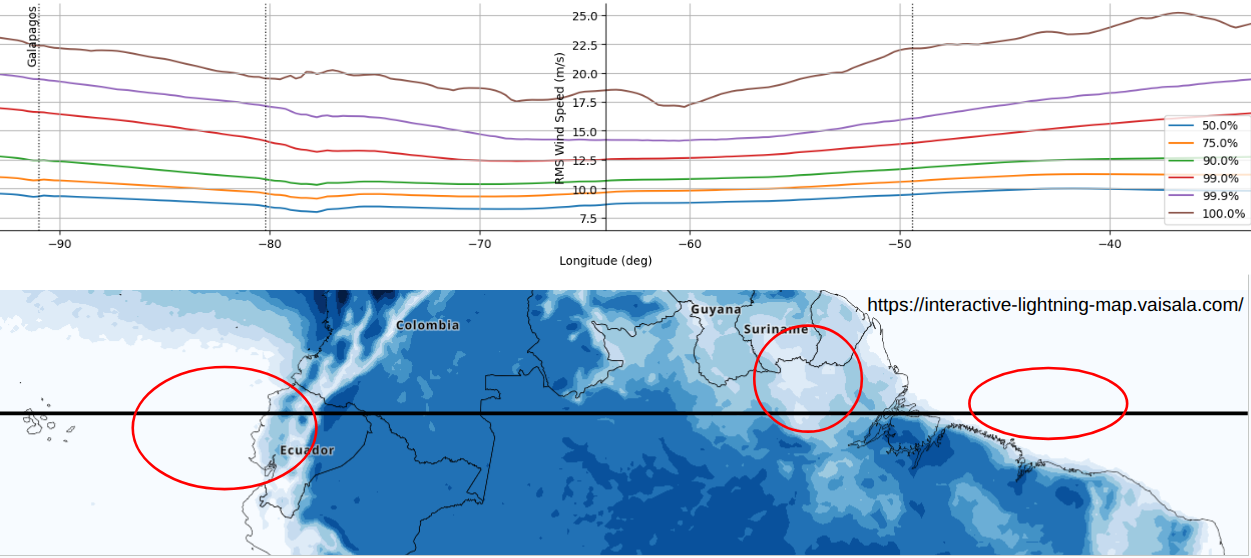}{
    America}
  \mygraphsubfigure{fig:lightning-africa}{width=\columnwidth}{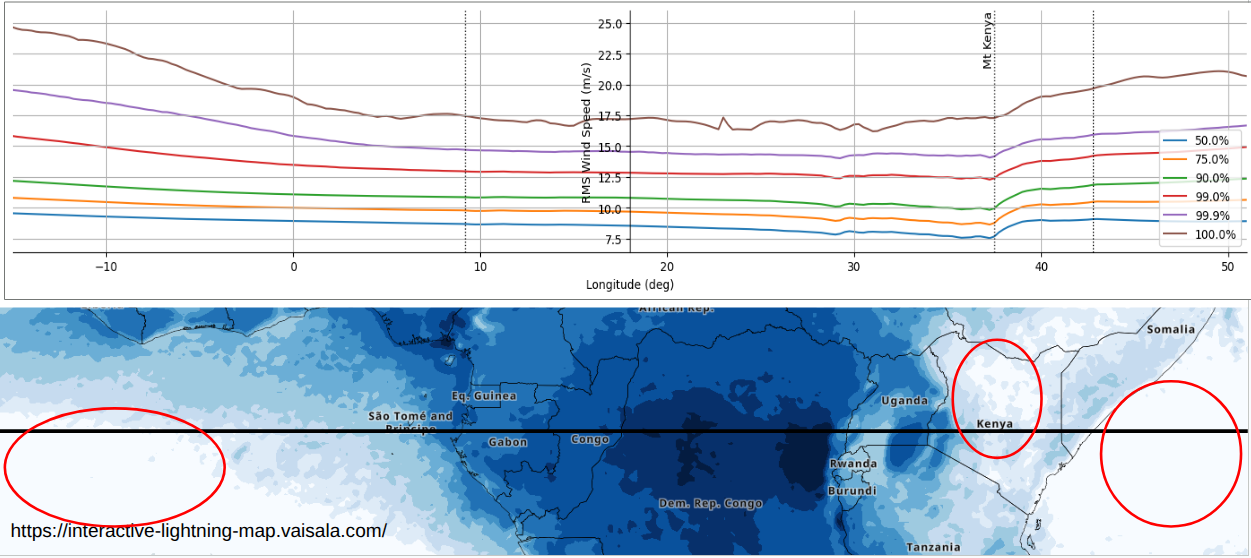}{
    Africa}
  \mygraphsubfigure{fig:lightning-indonesia}{width=\columnwidth}{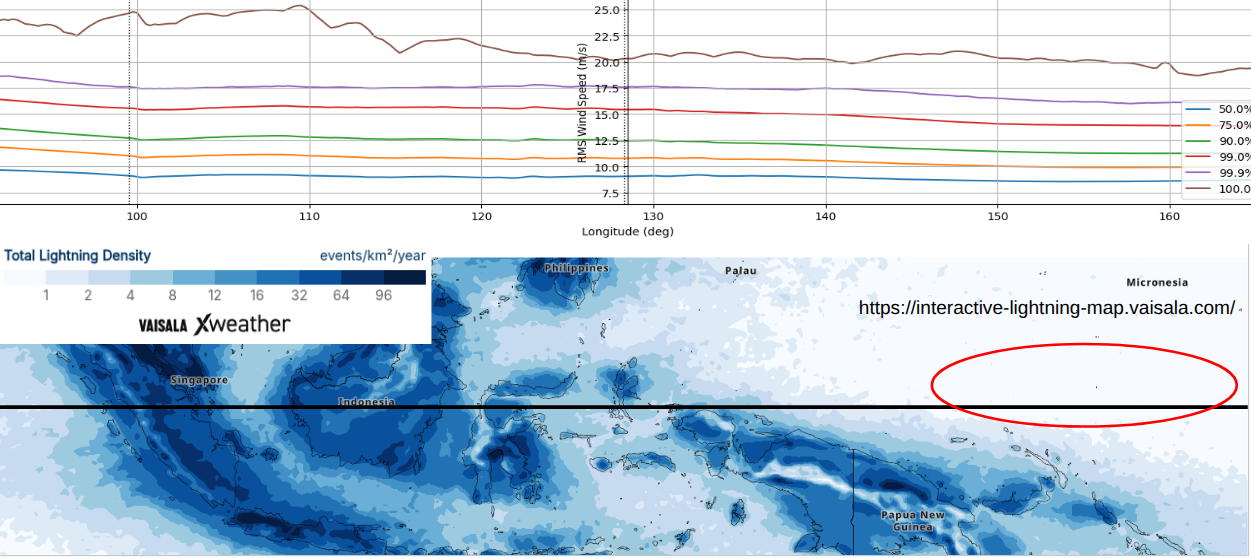}{
    Indonesia}
}{Comparing wind and lightning data suggests a few most promising locations
for space elevator deployment. Lightning data reproduced with permission from Vaisala
Xweather.}

\section{Elevator Design for Wind}
\label{sec:design}

With the results of the preceding sections, we can now write out the
requirements that need to be met by an elevator design. A survival spec
sets the highest wind load that can be tolerated. A launch spec sets a
lower wind load during which climber launch must be possible. Finally, a 
debris resistance spec constrains the width of the ribbon once it gets to
LEO. In addition to wind speed, the wind loading specs need to specify the
tension in the ribbon and a criterion for excessive wind impact, either
inclination or load fraction. Table~\ref{tab:se-spec} gives an example spec
which we will attempt to design for.

The reader will note that the launch spec indicates conditions both above
and below the climber. Indeed, there is a sharp drop in tension below the
climber, reducing $v_c$ and making the ribbon more susceptible to the wind.
Moreover, while load fraction is important above the climber, as it
determines how much mass can be launched, it is no longer important below
the climber and we can just set an inclination spec sufficient to ensure
that the climber can't be blown back down by the tether inclining under
high wind.

\mytable{tab:se-spec}{
\begin{tabular}{|l|}
\hline
\textbf{Survival Spec:}              \\
  - $\vrms < 30~\mathrm{m/s}$     \\
  - Tension: 20~kN               \\
  - Inclination: $<70\deg$       \\

\textbf{Launch Spec:} \\
  - $\vrms < 15~\mathrm{m/s}$ \\
  - Above the climber: \\
    \qquad - Tension: 200~kN         \\
    \qquad - Load fraction: 90~\%.   \\

  - Below the climber:            \\
    \qquad - Tension: 20~kN                 \\
    \qquad - Inclination: $<70\deg$        \\

\textbf{Debris Resistance Spec:}              \\
  - Wide tether above 150~km. \\

\hline
\end{tabular}
}{Example specs allowing a space elevator to be designed.}

With the specs in place we can use Figure~\ref{fig:design-chart} to set the
maximum width of the space elevator near the ground. This chart contains
the same information as Figures~\ref{fig:inclination}
and~\ref{fig:load-fraction}, presented in a way that allows values to
simply be looked up. To use it, draw a horizontal line through the desired load fraction 
or inclination limit, intersect that line with the curve corresponding to the
maximum $\vrms$, and draw a vertical line through the intersection point to
determine the maximum allowed tether width.

Repeating this exercise for the survival, above climber and below climber
specs, we find, for the example in Table~\ref{tab:se-spec}, maximum widths of
40~mm, 42~mm and 16~mm, respectively. Hence we will select a 16~mm-wide
ribbon. This 16~mm ribbon selected for wind robustness below the climber
will imply high compressive stresses on the climber rollers in order to get
sufficient traction. We shall see in the next sections ways to reduce this
stress.

\mygraphfigure{fig:design-chart}{width=\columnwidth}{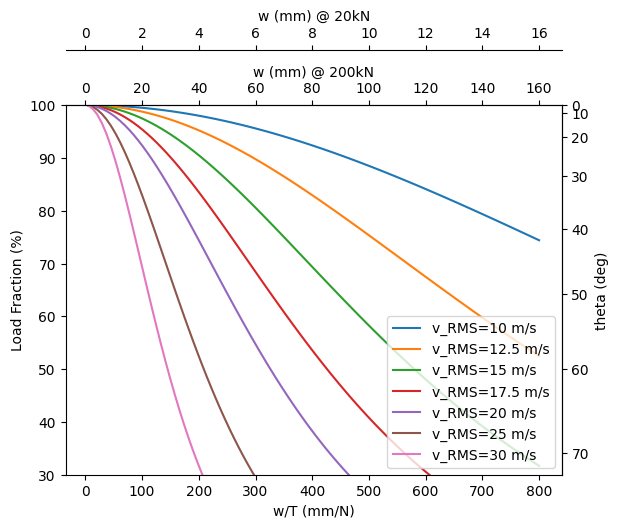}{Chart
showing the load fraction and inclination as a function of
width-over-Tension ratio for various wind speeds. For convenience the top
axes show the width assuming tensions of 20~kN or 200~kN.}

\subsection{Reel-Out Launch}

Reel-out launch, also known as spring-forward, is an approach first
introduced by~\cite{knapman2014design} based on an idea by Ben Shelef, in which
the climber is launched from Earth by reeling out ribbon from the
anchor, rather than by having the climber climb the ribbon. Among several
advantages, this approach allows the ribbon below the climber to be
narrower than the ribbon above the climber, since they are two different
pieces of material, an advantage already noted in~\cite{robinson2022}.

Continuing with our numerical
example, the ribbon below the climber still has to be 16~mm wide, but now
the ribbon above the climber can be 40~mm wide, or even 42~mm wide if we
are willing to make the operational commitment to reel out some of the 16~mm
ribbon in survival spec wind conditions. This more than 2.5 times increase
in ribbon width may facilitate climber roller design.

\myfigurewide{fig:climber-vs-reel-out-launch}{
\mygraphsubfigure{fig:climbing-launch}{width=0.98 \columnwidth}{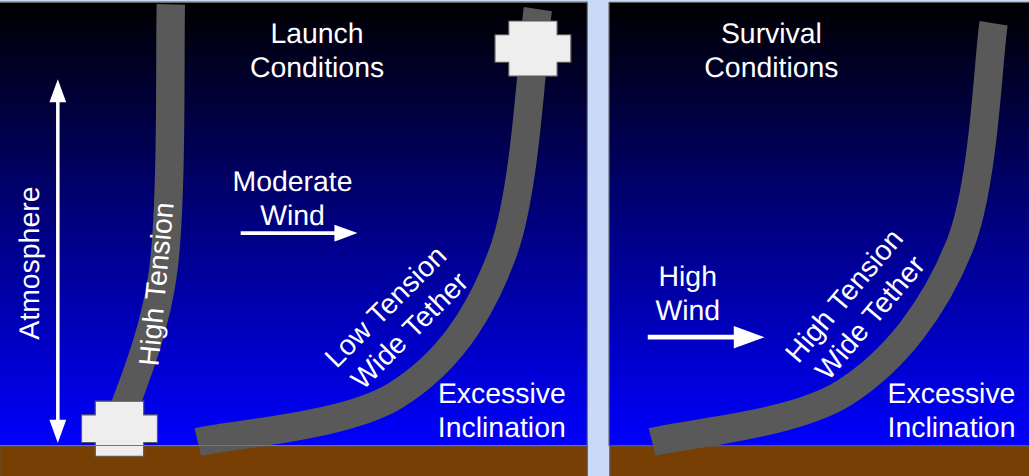}
{Climbing launch.}
\hfill
\mygraphsubfigure{fig:reel-out-launch}{width=0.98 \columnwidth}{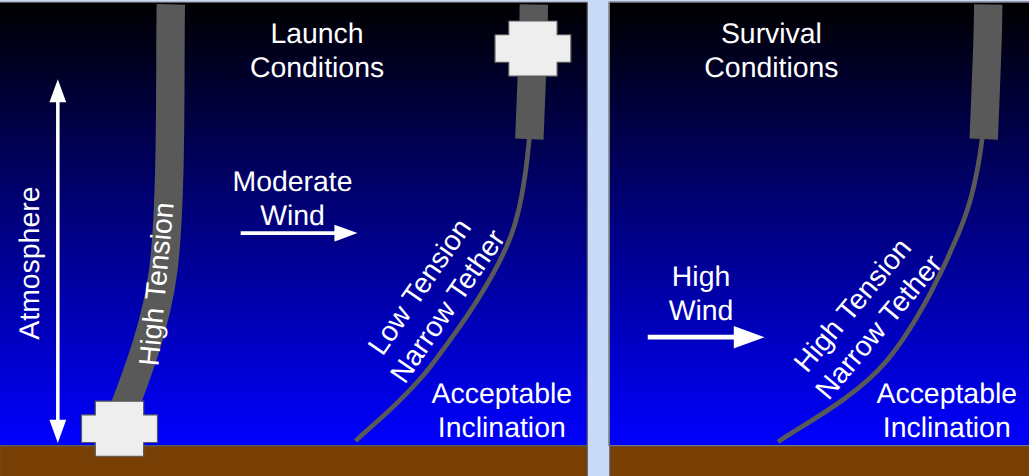}
{Reel-out launch.}
}{Comparison between a climbing launch~\ref{fig:climbing-launch} and a
reel-out launch~\ref{fig:reel-out-launch}. With the climbing launch, the 
wide elevator ribbon is vulnerable during high-wind events or when there is a
climber at low altitude. In both of these vulnerable scenarios the wind loading
can be reduced by reeling out a narrower, possibly non-climbable tether.}

\subsection{Capstan Drive}

Even with a 42~mm ribbon, we are well below what is considered the minimum
climbable width in~\cite{wright2023climber}. A key limitation is the
stress in the Ti-6Al-4V alloy rollers, which is close to 50~\% of the
material yield stress for a 310~mm wide ribbon. Shrinking the ribbon to
42~mm would increase the stress to an unacceptable 370~\% of yield.
                                                                       
This problem can be addressed by switching away from the pinch drive,
where the ribbon is pinched between pairs of rollers that are pressed
against each other, in favor of a capstan drive where traction is achieved
by having the ribbon alternatively wind left and right between offset
rollers as in Figure~\ref{fig:capstan}. With this configuration the contact
area on the rollers is increased, and the stress is reduced accordingly.

\mygraphfigure{fig:capstan}{width=\columnwidth}{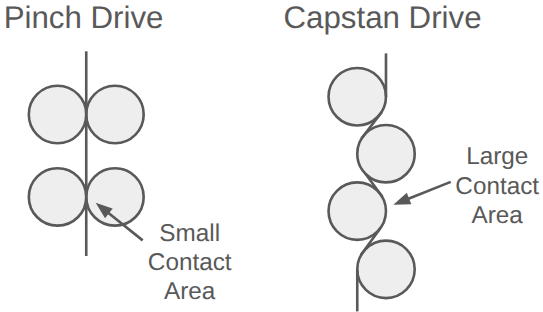}{The pinch
drive (left) achieves traction by pinching the ribbon between two rollers.
A larger contact area can be obtained using a capstan drive (right).}

The capstan drive is currently out of favor in the climber competition
community, possibly because it can suffer from slippage problems, though
we have yet to find references outlining the conditions in which
the slippage occurs. Reference~\cite{wright2023climber} considers the
capstan drive, but dismisses it because of the large transverse forces it
would entail. However, the transverse force needed to achieve the necessary
traction on a capstan drive is in fact a bit smaller than on a pinch drive
because part of the traction comes from surfaces that are non-vertical. It
will be necessary, though, with a capstan drive to let the spacing of the
rollers increase as the climber ascends to keep the transverse forces from
increasing as the tension in the ribbon increases with altitude. The
mechanical complexity of this adaptability should be comparable to the
complexity of maintaining a constant pressure on the ribbon in a pinch
drive, and may in fact be less challenging if there are imperfections in
the ribbon due to debris impact and patches that need to be accommodated.

Much more could be said about the capstan drive, but returning to the
context of wind loading, it seems like a way to address the necessary
narrowness of the tether in the atmosphere that is much simpler than the
approaches proposed in~\cite{robinson2022} such as Lofstrom
loops~\cite{lofstrom1985launch},
multi-staged elevators~\cite{knapman2019multistage}, Thoth
towers~\cite{quine2015spaceelevator} or transferring from a winch to a
climbable elevator outside the atmosphere.

\subsection{Kite Survival Mode}

A final trick that can be used for extreme wind events is to deploy a kite
onto the tether. The kite is attached at the base and reeled out. It
provides lift, increasing the tension in the tether, but reducing its
curvature. The kite also has drag which causes a kink in the ribbon.
Detailed analysis of this concept is left for future work.

\mygraphfigure{fig:kite}{width=\columnwidth}{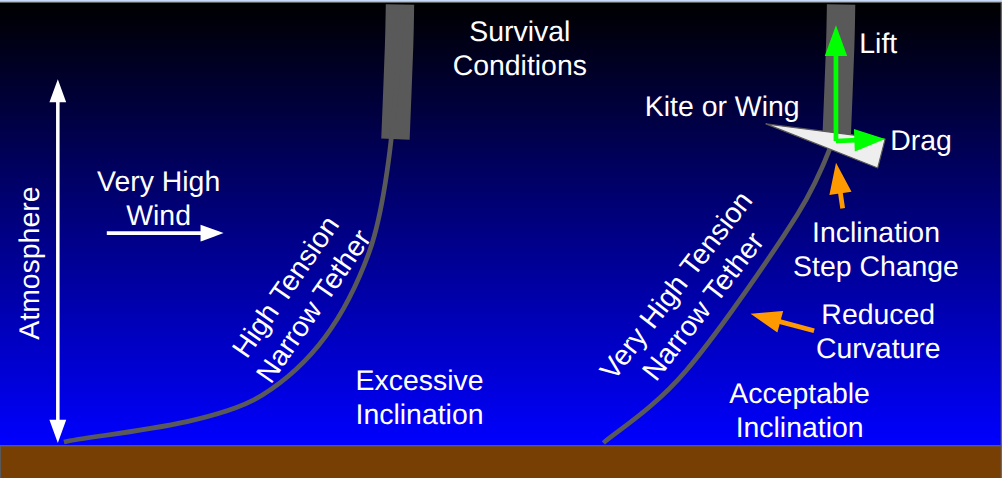}{During extreme wind events,
a kite could be raised to increase the tension in the ribbon reducing its
inclination.}

\subsection{Tether Alignment Relative to the Wind}

So far we have assumed that the ribbon is facing the wind. This is a
question that merits further study as an edge-on configuration could reduce
the drag coefficient by more than 100x, which would render the wind almost
negligible. Alternatively, the added degree of freedom provided by the
orientation of the ribbon could add challenges. The ribbon is long and has 
low stiffness in the transverse and twisting directions,
leaving open the possibility of aeroelastic effects. If flutter arises, it
could lead to increased drag because of lift caused by the
side-to-side motion, or lead to increased tension in the elevator requiring
a larger factor of safety.

One approach to controlling orientation relative to the wind could be to
add features to the edge of the ribbon breaking its symmetry and favoring
an edge-wise orientation. For instance, providing excess material on one
side of the ribbon at regular intervals could encourage the ribbon to
weather-vane to an edge on configuration. These features would have to be
designed to be compatible with the passage of climbers.

Figure~\ref{fig:wind-orientation} shows an example of wind orientation as a
function of altitude from ERA5 data.
There is a lot of variation of orientation with altitude so maintaining
a fixed orientation would require twisting the ribbon. But if it is too
thin, the ribbon could undergo transverse buckling when
twisted~\cite{chopin2013helicoids, gassend2025twist}, where the ribbon
folds rather than twists when its orientation varies. This could cause
ribbon damage, particularly if a climber was to crease the folded section
while climbing.
Even if no voluntary attempt is made to orient the ribbon
edge-on, the ribbon may spontaneously choose some alignment relative to the
wind, causing twist, which in turn could cause transverse buckling.

\mygraphfigure{fig:wind-orientation}{width=\columnwidth}{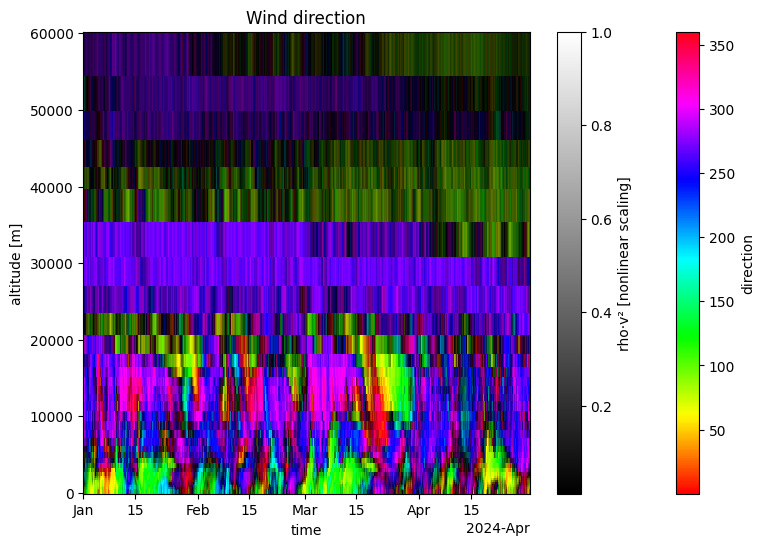}{Wind
as a function of time and altitude at 90\textdegree{} East longitude
during April 2024. Color shows heading (0\textdegree{} is North,
90\textdegree{} is East) and brightness shows dynamic pressure.}
                                                                                      
The ribbon orientation is an important and challenging question which will
have to be tackled to fully understand the impact of the wind on the space
elevator.

\section{Conclusion}

We have analyzed the impact of the wind on a space elevator, and shown that
under suitable approximations, the impact of the wind can be understood
from $\vrms$, the RMS wind velocity integrated over pressure. Under the effect of
the wind, the elevator inclines itself.  If the inclination of the elevator
gets too close to 90\textdegree{} a pull-down phenomenon can occur where the
elevator is blown away and eventually pulled down by the wind. We have
proposed that pull-down avoidance and maximization of payload are the two
criteria that should determine how much wind is too much for a given
elevator design.

This analytical work has allowed us to reduce the ERA5 atmospheric data to
a single plot showing the impact of wind on a space elevator as a function
of longitude. Notably, $\vrms$ was found to be highest in
the center of oceans (max \textless~30~m/s, 90\textsuperscript{th}
percentile \textless~17~m/s), and lowest over continents (max
\textgreater~16~m/s, 90\textsuperscript{th}
percentile \textgreater~12~m/s). Statistics on $Pv^2$ suggested that
the approximately 60~km maximum altitude of the ERA5 database misses a
modest amount of the wind loading. Some additional patterns were identified
in the ERA5 data: The vertical component is small and can be ignored. There
is substantial ocean-dependent seasonal variation.  There is also a lot of
year-to-year variation, in some cases dominated by a single large storm.

Combining ERA5 with lightning data from Vaisala Xweather suggested
promising space elevator locations: off the equatorial coasts of Ecuador,
Brazil and Somalia; South of Suriname; Kenya; on the Equator South of
Liberia; and North of the Solomon Islands.

Finally, we considered how to design a space elevator to be resistant to
the expected wind loads. The equatorial winds are too large for the
International Space Elevator Consortium's 30~cm wide reference design
tether to be able to survive, so several methods were proposed to improve
wind resistance: the reel-out method (also known as spring-forward), the
use of capstan drives, and for extreme conditions the use of kites. Adding
features to the ribbon that would align it edge-on to the wind was also
considered.

An important avenue of future investigation lies in going beyond the
assumption that the elevator ribbon faces the wind, as adding this degree
of freedom could either facilitate the space elevator's wind resistance or
make it more challenging. The methods to foster wind resistance also merit further
development. And finally, how climate change will affect the
historical data needs to be considered.

\section{Acknowledgements}

The results presented in this paper use data from the Copernicus Climate
Change Service (C3S). The results contain modified Copernicus Climate
Change Service information 2020. Neither the European Commission nor ECMWF
is responsible for any use that may be made of the Copernicus information
or data it contains.

Lightning maps that were presented are from Vaisala Xweather.

ChatGPT 5.2 and Gemini3 Pro were used as an aid in proof-reading and in
identifying sections that need further attention.

The authors would like to recognize the members of the International Space
Elevator Consortium's Powering the Space Elevator working group for
inspiring discussions and feedback, and in particular John Knapman, Larry
Bartozsek and Adrian Nixon.

Thanks finally to Dennis Wright for waiting well past his proceedings
deadline for this paper. The combination of time pressure and extra time were 
instrumental in getting the paper written.

\bibliographystyle{alpha}
\bibliography{paper}

\end{document}